\documentclass[aps,twocolumn,superscriptaddress,showpacs]{revtex4}%
\usepackage{amsfonts}
\usepackage{amsmath}
\usepackage{amssymb}
\usepackage{graphicx}%
\providecommand{\U}[1]{\protect\rule{.1in}{.1in}}
\begin{document}
\title{Many-body reduced fidelity susceptibility in Lipkin-Meshkov-Glick model}
\author{Jian Ma}
\affiliation{Department of Physics and ITP, The Chinese University of Hong Kong, Hong Kong, China}
\affiliation{Zhejiang Institute of Modern Physics, Department of Physics, Zhejiang
University, Hangzhou 310027, P. R. China.}
\author{Xiaoguang Wang}
\email{xgwang@zimp.zju.edu.cn}
\affiliation{Zhejiang Institute of Modern Physics, Department of Physics, Zhejiang
University, Hangzhou 310027, P. R. China.}
\author{Shi-Jian Gu}
\email{sjgu@phy.cuhk.edu.hk}
\affiliation{Department of Physics and ITP, The Chinese University of Hong Kong, Hong Kong, China}
\date{\today }

\pacs{64.70.Tg, 03.67.-a, 03.65.Ud, 75.10.Jm}

\begin{abstract}
We study the reduced fidelity susceptibility $\chi_{r}$ for an
$M$-body subsystem of an $N$-body Lipkin-Meshkov-Glick model with
$\tau=M/N$ fixed. The reduced fidelity susceptibility can be viewed
as the response of subsystem to a certain parameter. In noncritical
region, the inner correlation of the system is weak, and $\chi_{r}$
behaves similar with the global fidelity susceptibility $\chi_{g}$,
the ratio $\eta=\chi_{r}/\chi_{g}$ depends on $\tau$ but not $N$.
However, at the critical point, the inner correlation tends to be
divergent, then we find $\chi_{r}$ approaches $\chi_{g}$ with the
increasing the $N$, and $\eta=1$ in the thermodynamic limit. The
analytical predictions are perfect agreement with the numerical
results.

\end{abstract}
\maketitle

\section{Introduction}

Quantum phase transition (QPT) \cite{sachdev} occurs at absolutely zero
temperature is driven purely by quantum fluctuations. It was studied
conventionally by Landau paradigm with order parameter in the frame of
statistics and condensed matter physics. Recently, two quantum-information
\cite{nilesen} concepts, entanglement
\cite{xgwPRA64,vidal03,Latorre04,Osterloh,vidal04,Sebastien04,Latorre05,BarthelPRA74,BarthelPRL97,Roman08,Cui08}
and fidelity
\cite{HTQuan06,PZanardi06,Buonsante07,PZanardi0606130,PZanardi07,WLYou07,HQZhou07,LCVenuti07,SJGu07,SChen07,WQNing07,MFYang07,NPaunkovic07,KwokPRE78,MaPRE78}
have been investigated extensively in QPTs and are recognized to be effective
and powerful in detecting the critical point. The former measures quantum
correlations between partitions, while the latter measures the distance in
quantum state space. Therefore, the success of them in characterizing QPTs is
understood by regarding the universality of the critical behaviors itself,
that is, the divergent of the correlation and the dramatic change of the
ground state structure. Furthermore, as the fidelity depends computationally
on an arbitrarily small change of the driving parameter, Zarnardi \textit{et
al}. suggested the Riemannian metric tensor \cite{PZanardi07}, while You
\textit{et al}. suggested the fidelity susceptibility \cite{WLYou07}, both
focus on the leading term of the fidelity. In the following, we mainly
consider the fidelity susceptibility (FS).

Until now, most efforts have been devoted to the study of the global
ground state fidelity susceptibility (GFS), denoted by $\chi_{g}$,
which reflects the susceptibility of the system in response to the
change of certain driving parameter. In this work, we study the
responses of a subsystem, for which we study its FS, the so-called
reduced fidelity susceptibility (RFS), denoted by $\chi_{r}$. Some
special cases have been studied in Refs.
\cite{HQZhou07,NPaunkovic07,KwokPRE78,MaPRE78}, where the subsystems
are only one-body or two-body, while in this paper we will study an
arbitrary $M$-body subsystem. The motivation for the investigation
of RFS is clear in physics. Firstly, it reveals information about
the change of the inner structure for a system that undergoes QPT.
Secondly, as the existence of interactions and correlations, a
general quantum system is not the simple addition of its different
parts, especially in the critical region, where the entanglement
entropy is divergent \cite{vidal03,Latorre04,BarthelPRL97}.
Therefore it is significant to investigate the behavior of the RFS,
as well as the effects of entanglement on it, in both critical and
noncritical regions. And our study can be viewed as a connection
between the FS and the entanglement entropy.

To study this question, we consider an $N$-body Lipkin-Meshkov-Glick
model (LMG)~\cite{Lipkin} model, and study the RFS for its $M$-body
subsystem. As $0\leq\chi_{r}\leq\chi_{g}$ \cite{MaPRE78}, we
consider a more useful quantity, $\eta=\chi_{r}/\chi_{g}$, and thus
$\eta\in\lbrack0,1]$. We find that, the behaviors of the RFS, as
well as $\eta$, are quite different in noncritical and critical
regions. In noncritical region, the entanglement entropy is
saturated by a finite upper bound, and the inner correlation is
small, thus the RFS behaves similar with the GFS, and the ratio
$\eta$ depends on $\tau=M/N$ but not $N$. However, at the critical
point, the entanglement entropy tends to be divergent with the
increasing of system size, and the inner correlations are very
strong. Then we find the RFS approaches GFS with the increasing of
$N$, and $\eta=1$ in the thermodynamic limit for $\tau\neq 0$. These
can be understood by considering the divergent of correlation in
second-order QPTs, which is reflected by the entanglement entropy.

This paper is organized as follows. In Sec. II, we introduce the LMG model and
give a brief review of the GFS studied in \cite{KwokPRE78}. Then in Sec. III,
we derive the RFS in the thermodynamic limit and obtain its divergent form in
the vicinity of the critical point. Then we perform some numerical
computations, and the results are in perfect agreement with our analytical prediction.

\section{LMG model and global fidelity susceptibility}

The LMG model, originally introduced in nuclear physics and has
found applications in a broad range of other topics: statistical
mechanics of quantum spin system \cite{BotetPRL49}, Bose-Einstein
condensates \cite{Cirac}, or magnetic molecules such as Mn$_{12}$
acetate \cite{Garanin}, as well as quantum entanglement
\cite{VidalPRA69}, and quantum fidelity \cite{KwokPRE78,MaPRE78}. It
is an exactly solvable \cite{PanPLB451,LinksPRA36}\ many-body
interacting quantum system as well as one of the simplest to show a
quantum transition in the regime of strong coupling. The quantum
phase transition of this model can be described by the symmetry
broken mechanism, the two phases are associated with either
collective or single-particle behavior. The Hamiltonian of the LMG
model reads%
\begin{equation}
H=-\frac{1}{N}\left(  S_{x}^{2}+\gamma S_{y}^{2}\right)  -hS_{z},
\end{equation}

where $S_{\alpha}=\sum_{i=1}^{N}\sigma_{\alpha}^{i}/2$
($\alpha=x,y,z$) are the collective spin operators;
$\sigma_{\alpha}^{i}$ are the Pauli matrices; $N$ is the total spin
number; $\gamma$ is the anisotropic parameter. $\lambda$ and $h$ are
the spin-spin interaction strength and the effective external field,
respectively. Here, we focus on the ferromagnetic case
($\lambda>0$), and without loss of generality, we set $\lambda=1$
and $0\leq\gamma\leq1$. As the spectrum is invariant under the
transformation $h\leftrightarrow-h$, we only consider $h\geq0$. This
system undergoes a second-order QPT at $h=1$, between a symmetric
(polarized, $h>1$) phase and a broken (collective, $h<1$) phase,
which is well described by a mean-field approach \cite{DusuelPRB71}.
The classical state is fully polarized in the field direction
$\left(  \left\langle \sigma_{z}^{i}\right\rangle =1\right)  $
for $h>1$, and is twofold degenerate with $\left\langle \sigma_{z}%
^{i}\right\rangle =h$ for $h<1$.

Before deriving the RFS, we give a brief review of the GFS of the
LMG model that has been studied in Ref. \cite{KwokPRE78}, where the
authors employed the Holstein-Primakoff transformation and derived
the GFS for both phases in the thermodynamic limit,\begin{widetext}
\begin{equation}
\chi_{g}\left(  h,\gamma\right)  =\left\{
\begin{aligned} &\frac{N}{4\sqrt{\left( 1-h^{2}\right) \left( 1-\gamma\right) }}
+\frac{h^{2}\left( h^{2}-\gamma\right) ^{2}}{32\left(
1-\gamma\right) ^{2}\left( 1-h^{2}\right) ^{2}},
&\text{for}\quad&0\leq h<1,\\ &\frac{\left( 1-\gamma\right)
^{2}}{32\left( h-\gamma\right) ^{2}\left( h-1\right) ^{2}},
&\text{for}\quad& h \ge 1. \end{aligned}\right. \label{gfs}
\end{equation}
\end{widetext}It has been found that, when $h<1$, the GFS increases with $N$
and can be viewed as an extensive quantity, however, when $h>1$ the GFS is
saturated with an upper bound, i.e. it is intensive.

\section{Reduced fidelity susceptibility}

\subsection{Thermodynamic limit}

Now we give some basic formulas for fidelity and its susceptibility. As the
subsystem is represented by a mixed state, we introduce the Uhlmann fidelity
\cite{Uhlmann},
\begin{equation}
F\left(  \rho,\tilde{\rho}\right)  \equiv \text{tr}\sqrt{\rho^{1/2}\tilde{\rho}%
\rho^{1/2}},
\end{equation}
where $\rho\equiv\rho\left(  h\right)  $ and $\tilde{\rho}\equiv\rho\left(
h+dh\right)  $ with a certain parameter $h$. If $dh$ tends to zero, the two
states are close in parameter space, and their Bures distance \cite{Bures69}
is,%
\begin{equation}
ds_{B}^{2}=2\left[  1-F\left(  \rho,\tilde{\rho}\right)  \right]  .
\end{equation}
In the basis of $\rho$, denoted by $\left\{
|\psi_{i}\rangle\right\} $, the
Bures distance can be written as \cite{SommerJPA}%
\begin{equation}
ds_{B}^{2}=\frac{1}{4}\sum_{n=1}^{N}\frac{dp_{n}^{2}}{p_{n}}+\frac{1}{2}%
\sum_{n\neq m}^{N}\frac{\left(  p_{n}-p_{m}\right)  ^{2}}{p_{n}+p_{m}%
}\left\vert \langle\psi_{n}|d\psi_{m}\rangle\right\vert ^{2},
\end{equation}
where $p_{i}$ are the eigenvalues of $\rho$, $N$ is the dimension of
$\rho$. As FS is the leading term of fidelity, i.e.,
$F=1-\chi\delta^{2}/2$, we can get
FS for $h$ immediately,%
\begin{equation}
\chi\left(  h\right)  =\frac{1}{4}\sum_{n=1}^{N}\frac{\left(  \partial
_{h}p_{n}\right)  ^{2}}{p_{n}}+\frac{1}{2}\sum_{n\neq m}^{N}\frac{\left(
p_{n}-p_{m}\right)  ^{2}}{p_{n}+p_{m}}\left\vert \langle\psi_{n}|\partial
_{h}\psi_{m}\rangle\right\vert ^{2}, \label{chi}%
\end{equation}
where $\partial_{h}:=\partial/\partial h$. In our study, $\rho$ and
$\tilde{\rho}$ are just the reduced density matrices for ground states.

In the follows, the $N$-body LMG is divided into two parts, $A$ and
$B$ with size $M$ and $N-M$, respectively. We will study the RFS for
subsystem $A$, the reduced density matrix is $\rho_{A}$. This study
would give a connection between the RFS and the entanglement entropy
\cite{BarthelPRL97}. As we know that, the entanglement reflects the
correlation among inner partitions, and our study will reveal the
effects of these correlations on RFS, especially at the critical
point.

Now we introduce the total spin operators for the two subsystems,
$S_{\alpha}^{A,B}=\sum_{i\in A,B}\sigma_{\alpha}^{i}/2$. To describe
quantum fluctuations, it is convenient to use the Holstein-Primakoff
representation of the spin operators \cite{HolsteinPR58}, and the
first step is to rotate the
$z$ axis along the semiclassical magnetization%
\begin{equation}%
\begin{pmatrix}
S_{x}\\
S_{y}\\
S_{z}%
\end{pmatrix}
=%
\begin{pmatrix}
\cos\theta_{0} & 0 & \sin\theta_{0}\\
0 & 1 & 0\\
-\sin\theta_{0} & 0 & \cos\theta_{0}%
\end{pmatrix}%
\begin{pmatrix}
\tilde{S}_{x}\\
\tilde{S}_{y}\\
\tilde{S}_{z}%
\end{pmatrix}
.\label{rotate}%
\end{equation}
As presented in \cite{DusuelPRB71}, $\theta_{0}=0$ for $h>1$ so that
$\mathbf{S}=\mathbf{\tilde{S}}$, and $\theta_{0}=\arccos h$ for $h\leq1$. The
Holstein-Primakoff representation is then applied to the rotated spin
operators%
\begin{align}
\tilde{S}_{z}^{A} &  =M/2-a^{\dagger}a,\nonumber\\
\tilde{S}_{-}^{A} &  =\sqrt{M}a^{\dagger}\sqrt{1-a^{\dagger}a/M}=\left(
\tilde{S}_{+}^{A}\right)  ^{\dagger},\nonumber\\
\tilde{S}_{z}^{B} &  =\left(  N-M\right)  /2-b^{\dagger}b,\nonumber\\
\tilde{S}_{-}^{B} &  =\sqrt{N-M}b^{\dagger}\sqrt{1-b^{\dagger}b/\left(
N-M\right)  }=\left(  \tilde{S}_{+}^{B}\right)  ^{\dagger},
\end{align}
where $a\left(  a^{\dagger}\right)  $ and $b\left(
b^{\dagger}\right)  $ are bosonic creation and annihilation
operators for subsystem $A$ and $B$, respectively, and
$S_{\pm}^{A,B}=S_{x}^{A,B}\pm iS_{y}^{A,B}$. After this
transformation, the LMG Hamiltonian is mapped onto a system of two
interacting bosonic modes $a$ and $b$. For fixed $\tau=M/N$, the
Hamiltonian can be expanded in $1/N$. Up to the order $\left(
1/N\right) ^{0}$, one gets $H=NH^{(-1)}+H^{\left(  0\right)
}+O\left( 1/N\right)  $ with $H^{\left(
-1\right)  }=(m^{2}-1-2h)/4$, where $m=\cos\theta_{0}$, and%
\begin{equation}
H^{(0)}=-\frac{1+\gamma}{4}+\mathbf{A}^{\dagger}\mathbf{VA}^{T}+\frac{1}%
{2}\left[  \mathbf{A}^{\dagger}\mathbf{W}\left(  \mathbf{A}^{\dagger}\right)
^{T}+h.c.\right]  \label{bosonic_ham}%
\end{equation}
where $\mathbf{A=}\left(  a,b\right)  $, and%
\begin{align}
\mathbf{V} &  \mathbf{=}\frac{2hm+2-3m^{2}-\gamma}{2}\mathbb{I}\nonumber\\
\mathbf{W} &  \mathbf{=}\frac{\gamma-m^{2}}{2}%
\begin{pmatrix}
\tau & \sqrt{\tau\left(  1-\tau\right)  }\\
\sqrt{\tau\left(  1-\tau\right)  } & 1-\tau
\end{pmatrix}
,
\end{align}
where $\mathbb{I}$ is a $2\times2$ identity matrix; $m=h$ in broken
phase and $m=1$ in symmetric phase. The bosonic Hamiltonian can be
diagonalized by Bogoliubov transformation and is useful in deriving
the reduced density matrix. As shown in
\cite{bombelli,PreschelJPA32,PeschelJPA36}, the reduced density
matrix for eigenstates of a quadratic form can always be written as
$\rho_{A}=e^{-\mathcal{H}}$ with%
\begin{equation}
\mathcal{H}=\kappa_{0}+\kappa_{1}a^{\dagger}a+\kappa_{2}\left(  a^{\dagger
2}+a^{2}\right)  .
\end{equation}
$\kappa_{i}$ ($i=0,1,2$) can be determined by using \cite{BarthelPRL97}%
\begin{equation}
\text{tr}\rho_{A}=1,~\text{tr}\left(  \rho_{A}a^{\dagger}a\right)
=\left\langle a^{\dagger}a\right\rangle ~\text{and}~\text{tr}\left(  \rho
_{A}a^{\dagger2}\right)  =\left\langle a^{\dagger2}\right\rangle .
\end{equation}
where $\left\langle \Omega\right\rangle =\left\langle
\psi_{g}|\Omega|\psi _{g}\right\rangle $, $|\psi_{g}\rangle$ is the
ground state, Then we can diagonalize $\rho_{A}$ by Bogoliubov
transformation. However, in this paper we will adopt another method
to diagonalize $\rho_{A}$, as shown in Ref. \cite{BarthelPRA74},
$\rho_{A}$ is written in the bosonic coherent
state representation%
\begin{align*}
\langle\phi|\rho_{A}|\phi^{\prime}\rangle &  =K\exp\left[  \frac{1}{4}\left(
\phi^{\ast}+\phi^{\prime}\right)  \frac{G^{++}-1}{G^{++}+1}\left(  \phi^{\ast
}+\phi^{\prime}\right)  \right]  \\
&  \times\exp\left[  \frac{1}{4}\left(  \phi^{\ast}-\phi^{\prime}\right)
\frac{G^{--}+1}{G^{--}-1}\left(  \phi^{\ast}-\phi^{\prime}\right)  \right]  ,
\end{align*}
where $a|\phi\rangle=\phi|\phi\rangle$; $K=\sqrt{\left(  1+G^{++}\right)
\left(  1-G^{--}\right)  }$ is determined by the normalization of $\rho_{A}$;
$G^{++}$ and $G^{--}$ are Green's functions defined as%
\begin{align}
G^{++} &  =\langle\left(  a^{\dagger}+a\right)  ^{2}\rangle,\nonumber\\
G^{--} &  =\langle\left(  a^{\dagger}-a\right)  ^{2}\rangle.
\end{align}
Then $\rho^{A}$ can be diagonalized by the following Bogoliubov
transformation,
\begin{align}
g &  =\cosh\varphi a+\sinh\varphi a^{\dagger}\nonumber\\
&  =\frac{P+Q}{2}a+\frac{P-Q}{2}a^{\dagger}%
\end{align}
with $PQ=1$, $PG^{++}=\mu Q$, and $QG^{--}=-\mu P$. The Green's functions can
be obtained by diagonalizing the bosonic represented Hamiltonian
(\ref{bosonic_ham}),%
\begin{align}
G^{++} &  =1+\left(  1/\alpha-1\right)  \tau,\nonumber\\
G^{--} &  =\left(  1-\alpha\right)  \tau-1,
\end{align}
where%
\begin{equation}
\alpha=\left\{
\begin{aligned} &\sqrt{\frac{h-1}{h-\gamma}} &\text{for}\quad&h\ge1,\\ &\sqrt{\frac{1-h^{2}}{1-\gamma}} &\text{for}\quad&0\le h<1. \end{aligned}\right.
\end{equation}
The diagonalized $\rho^{A}$ reads%
\begin{equation}
\rho^{A}=\frac{2}{\mu+1}e^{-\varepsilon g^{\dagger}g},
\end{equation}
where the pseudoenergy $\varepsilon=\ln\left[  \left(  \mu+1\right)  /\left(
\mu-1\right)  \right]  $ with $\mu=\alpha^{-1/2}\sqrt{\left[  \tau
\alpha+\left(  1-\tau\right)  \right]  \left[  \tau+\alpha\left(
1-\tau\right)  \right]  }$.

Now we can derive the RFS, of which the first term involves only the
eigenvalues of $\rho_{A}$, and the second term involves both the
eigenvalues and the eigenvectors. The eigenvectors of $\rho_{A}$ is
the number state $|n\rangle$: $g^{\dagger}g|n\rangle=n|n\rangle$,
and the term $\left\vert
\langle\psi_{n}|\partial_{h}\psi_{m}\rangle\right\vert
^{2}=\left\vert \langle
n|\partial_{h}m\rangle\right\vert ^{2}$ can be calculated by using%
\begin{equation}
\left\vert \langle n|\partial_{h}m\rangle\right\vert ^{2}=\frac{\left\vert
\left\langle n|\partial_{h}g^{\dagger}g|m\right\rangle \right\vert ^{2}%
}{\left(  m-n\right)  ^{2}}.
\end{equation}
Then we write the RFS explicitly,%
\begin{equation}
\chi_{r}\left(  h,\gamma,\tau\right)  =\frac{\left(  \partial_{h}\mu\right)
^{2}}{4\left(  \mu^{2}-1\right)  }+\frac{\left(  \mu\partial_{h}%
\varphi\right)  ^{2}}{\mu^{2}+1}+\frac{N\tau}{4\mu}\left(  \partial_{h}%
\theta_{0}\exp\varphi\right)  ^{2}, \label{rfs}%
\end{equation}
where $\varphi=\text{arctanh}\left[  \left(  \mu-G^{++}\right)
/\left( \mu+G^{++}\right)  \right]  $, $\theta_{0}=\arccos h$ for
$h\leq1$ and $\theta_{0}\equiv0$ for $h>1$. Thus the last term of
the above expression only takes effect in the broken phase. We
emphasize that, in the broken phase
$h<1$, we should perform a rotation (\ref{rotate}) at first.%
%TCIMACRO{\FRAME{ftbpFU}{2.8277in}{2.2254in}{0pt}{\Qcb{RFS as a function of $h$
%at $\gamma=1/2$ and $\tau=1/2$. The peaks approach to the critical point and
%become sharper and sharper with the increasing of $N$.}}{\Qlb{fig1}}%
%{fig1.eps}{\special{ language "Scientific Word";  type "GRAPHIC";
%width 2.8277in;  height 2.2254in;  depth 0pt;  original-width 6.9998in;
%original-height 4.9926in;  cropleft "0";  croptop "1";  cropright "1";
%cropbottom "0";  filename 'fig1.eps';file-properties "XNPEU";}} }%
%BeginExpansion
\begin{figure}
[ptb]
\begin{center}
\includegraphics[
height=2.2254in,
width=2.8277in
]%
{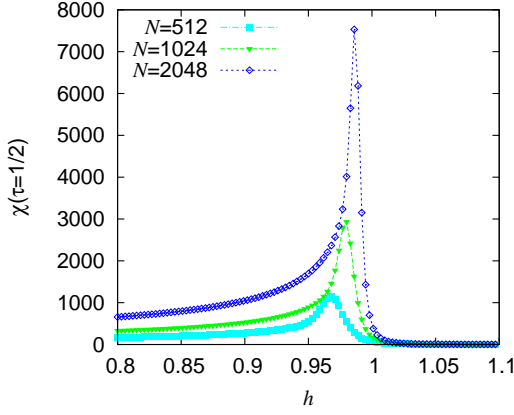}%
\caption{RFS as a function of $h$ at $\gamma=1/2$ and $\tau=1/2$.
The peaks approach the critical point and become sharper and sharper
with the
increasing of $N$.}%
\label{fig1}%
\end{center}
\end{figure}
%EndExpansion

We can express it farther as%
\begin{equation}
\chi_{r}\left(  h,\gamma,\tau\right)  =\left\{
\begin{aligned} &\chi+\frac{N\tau}{4G^{++}\left( 1-h^{2}\right) } &\text{for}\quad&0\leq h<1,\\ &\chi &\text{for}\quad& h \ge 1, \end{aligned}\right.
\end{equation}
where%
\begin{equation}
\chi=\frac{\left(  \partial_{h}\mu\right)  ^{2}}{4\left(  \mu^{2}-1\right)
}+\frac{\mu^{2}}{4\left(  \mu^{2}+1\right)  }\left[  \partial_{h}\ln\left(
-\frac{\mu}{G^{++}}\right)  \right]  ^{2}.
\end{equation}
In the vicinity of the critical point, the RFS diverges as%
\begin{align}
\chi_{r}/N\propto\left(  1-h\right)  ^{-1/2}\text{, }  &  \text{for }0\leq
h<1,\\
\chi_{r}\propto\left(  1-h\right)  ^{-2}\text{, }  &  \text{for }h\geq1,
\end{align}
and this is the same with $\chi_{g}$. Additionally, we show the
entanglement entropy $\mathcal{E}=-$tr$\left(  \rho\ln\rho\right)  $
that was derived in \cite{BarthelPRL97,BarthelPRA74},
\begin{equation}
\mathcal{E}=\frac{\mu+1}{2}\ln\frac{\mu+1}{2}-\frac{\mu-1}{2}\ln\frac{\mu
-1}{2}+x\ln2.
\end{equation}
where $x=1$ when $h<1$ and $x=0$ when $h>1$, the $\ln2$ term comes from the
two-fold degeneracy of the ground state in the broken phase, and this
degeneracy is lifted for finite $N$. The entanglement entropy diverges as
$\left(  1/4\right)  \ln\left\vert h-1\right\vert $ around the critical point,
and is nearly independent with $N$ in noncritical region.%
%TCIMACRO{\FRAME{ftbphFU}{2.8241in}{4.0287in}{0pt}{\Qcb{A comparison between
%$\eta$ (a) and $\QTR{cal}{E}$ (b) as a function of $h$ at $\gamma=1/2$,
%$\tau=1/2$ for various system sizes. At the critical point, $\eta$ tends to 1
%while $\QTR{cal}{E}$ is divergent.}}{\Qlb{fig2}}{fig2.eps}%
%width 2.8241in;  height 4.0287in;  depth 0pt;  original-width 6.992in;
%original-height 10.0024in;  cropleft "0";  croptop "1";  cropright "1";
%cropbottom "0";  filename 'fig2.eps';file-properties "XNPEU";}} }%
%BeginExpansion
\begin{figure}
[ptbh]
\begin{center}
\includegraphics[
height=4.0287in,
width=2.8241in
]%
{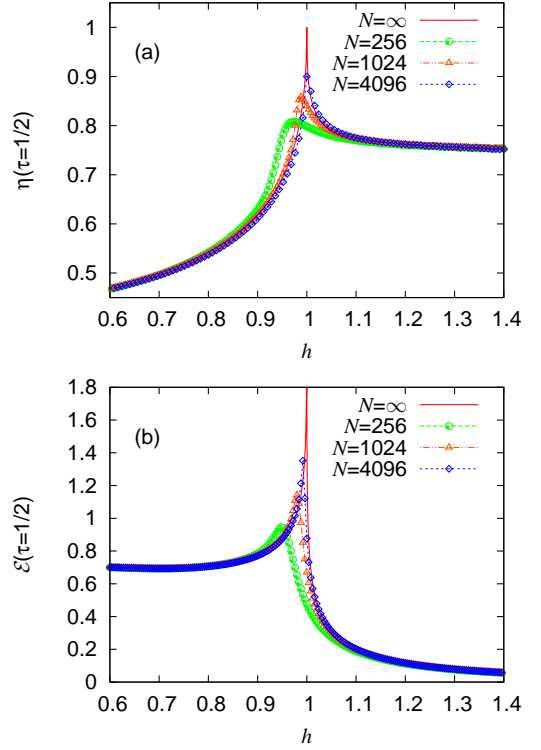}%
\caption{A comparison between $\eta$ (a) and $\mathcal{E}$ (b) as a function
of $h$ at $\gamma=1/2$, $\tau=1/2$ for various system sizes. At the critical
point, $\eta$ tends to 1 while $\mathcal{E}$ is divergent.}%
\label{fig2}%
\end{center}
\end{figure}
%EndExpansion

\subsection{Finite size cases}

To perform numerical computations, we should derive the reduced density matrix
for $\rho_{A}$ in finite size case. The LMG model is of high symmetry in
interaction, and the ground state which is the superposition of the Dick
states lies in the $J=N/2$ section%

\begin{equation}
|\psi_{g}\rangle=\sum_{m=0}^{N}C_{m}|J,-J+m\rangle,
\end{equation}
where $C_{m}$ is the coefficient to be determined numerically. We
hope to write $|J,-J+m\rangle$ in the form of
$|J_{A},m_{A}\rangle|J_{B},m_{B}\rangle $, where $J_{A}=M/2$ and
$J_{B}=\left(  N-M\right)  /2$ correspond to the two local systems.
Since $|J,-J+m\rangle=\sqrt{\left(  2J-m\right)  !/\left( 2J\right)
!m!}\left(  S_{+}\right)  ^{m}|J,-J\rangle$, and the ladder
operator $S_{+}=S_{+}^{A}+S_{-}^{B}$ . Then the ground state is%

\begin{align}
|\psi_{g}\rangle=  &  \sum_{m=0}^{N}\sum_{p=0}^{2J_{A}}C_{m}\sqrt
{\text{H}\left(  p;2J,2J_{A},m\right)  }|J_{A},-J_{A}+p\rangle\nonumber\\
&  \otimes|J_{B},-J_{B}+m-p\rangle\label{reducing of ground state}%
\end{align}
where%

\begin{equation}
\text{H}\left(  p;2j,2j_{1},m\right)  =\frac{\binom{2j_{1}}{p}\binom{2j_{2}%
}{m-p}}{\binom{2j}{m}} \label{hypergeometric distrbution}%
\end{equation}
is the so called Hypergeometric distribution function. And the matrix element
of $\rho_{A}$ is%

\begin{align}
\left(  \rho_{A}\right)  _{p,q}=  &  \sum_{m=0}^{N}C_{m}C_{q+m-p}^{\ast}%
\sqrt{\text{H}\left(  p;2J,2J_{A},m\right)  }\nonumber\\
&  \times\sqrt{\text{H}\left(  q;2J,2J_{A},q+m-p\right)  }.
\end{align}

By using the exact diagonalization method, the RFS as a function of
$h$ for fixed $\tau$ is computed and shown in Fig.~(\ref{fig1}). As
one can see that, the peaks of the RFS approach the critical point
and become sharper and sharper with the increasing of $N$. The RFS
in the symmetric phase ($h>1$) has an upper bound, however, in the
broken phase ($h<1$) the RFS increases with the total spin number
$N$. Thus we address that, the RFS is extensive in the broken phase,
in which the LMG model is of collective behavior, while is intensive
in the symmetric phase, in which the LMG model behaves like a single
particle. This is similar with the GFS \cite{KwokPRE78}.

\begin{figure}
[ptb]
\begin{center}
\includegraphics[
height=3.226in,
width=3.6677in
]%
{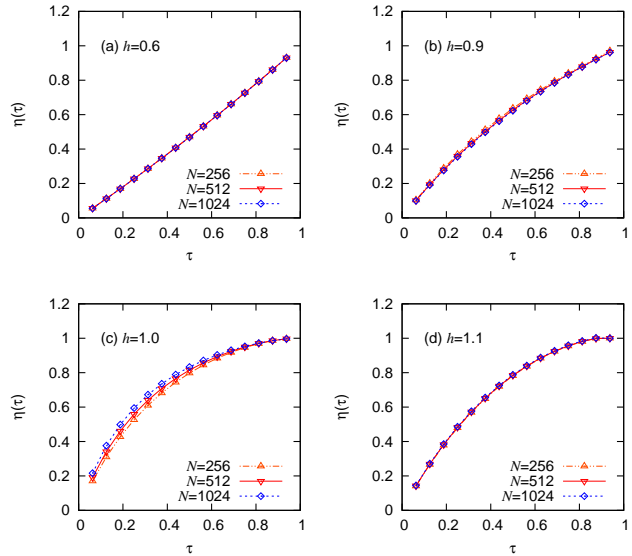}%
\caption{$\eta$ as a function of $\tau$ with $\gamma=1/2$, at
$h=0.6$ (a), $0.9$ (b), $1.0$ (c) and $1.1$ (d). We see that $\eta$
is nearly independent of $N$
when $h$ is away form the critical region.}%
\label{fig3}%
\end{center}
\end{figure}

As $0\leq\chi_{r}\leq\chi_{g}$, we will focus on a more useful quantity
$\eta\left(  \tau,h\right)  \equiv\chi_{r}\left(  h,\gamma,\tau\right)
/\chi_{g}\left(  h,\gamma\right)  $ and study its properties in critical and
noncritical regions. With Eqs.~(\ref{gfs}), (\ref{rfs}), we find that in the
thermodynamic limit%
\begin{equation}
\lim_{h\rightarrow1}\eta\left(  \tau,h\right)  =1,
\end{equation}
for any non-vanishing $\tau$. To verify our prediction, we show the
analytical and numerical results in Fig.~(\ref{fig2}). As one can
see that, at the critical point, the RFS approaches the global one,
i.e. $\eta$ tends to $1$, and at the same time, the entanglement
entropy, i.e. the inner correlation between subsystems $A$ and $B$,
is divergent with the increasing of $N$. When $h$ is away from the
critical region, the inner correlation decreases dramatically, and
then $\eta$ depends on $\tau$ but not the total system size $N$ as
shown in Fig.~(\ref{fig3}).
%TCIMACRO{\FRAME{ftbpFU}{3.6677in}{3.226in}{0pt}{\Qcb{$\eta$ as a function of
%$\tau$ with $\gamma=1/2$, at $h=0.6$ (a), $0.9$ (b), $1.0$ (c) and $1.1$ (d).
%We see that $\eta$ is independent of $N$ when $h$ is away form the critical
%region.}}{\Qlb{fig3}}{fig3.eps}{\special{ language "Scientific Word";
%type "GRAPHIC";  maintain-aspect-ratio TRUE;  display "USEDEF";
%valid_file "F";  width 3.6677in;  height 3.226in;  depth 0pt;
%original-width 9.0978in;  original-height 7.9952in;  cropleft "0";
%filename 'fig3.eps';file-properties "XNPEU";}} }%
%BeginExpansion
%EndExpansion

As demonstrated in Ref. \cite{MaPRE78}, when there are no correlations between
partitions of a system, for example an $N$-body system represented by a
product state that reads
\begin{equation}
|\psi\left(  h\right)  \rangle=\bigotimes_{i=1}^{N}|\phi_{i}\left(  h\right)
\rangle,
\end{equation}
if we denote a one-body reduced fidelity as $F_{r}$, the relation between the
global and the reduced fidelities is%
\begin{equation}
F_{g}\left(  h,\delta\right)  =\prod_{i=1}^{n}F_{r}^{i}\left(  h,\delta
\right)  .
\end{equation}
and thus we have $\chi_{g}=\sum_{i=1}^{N}\chi_{r}^{i}$, moreover, if
the system is of translation symmetry, we have $\chi_{g}=N\chi_{r}$.
If there is entanglement between partitions, we have no such
results, especially in the critical point, the entanglement is
divergent, and then $\chi_{g}/\chi_{r}=1$ in the thermodynamic
limit. This is some kind of effect of the inner correlations on the
susceptibility of the system states. However, we address that our
results are based on a high-dimension model, actually there are
interactions between any two particles in the LMG model. We think it
is deserved to study the RFS for a contiguous block in a
low-dimension model, for example, the $XY$ model in which the
interaction is just between neighboring sites. Thus the correlation
between a block and its complementary part takes effect only on the
boundary, and the results for $\eta$ maybe different.

\section{Conclusion}

In conclusion, we derive the RFS analytically in the thermodynamic
limit for a fixed $\tau$. To analyze the effects of the inner
correlations on the RFS, we study the ratio $\eta=\chi_{r}/\chi_{g}$
combined with the entanglement entropy in both critical and
noncritical regions. Our results give a clear picture for
understanding the effects of correlations on the response. In the
critical region, with the increasing of $N$, the entanglement
entropy tends to be divergent and $\eta$ approaches $1$, while in
the thermodynamic limit, $\eta\equiv1$ for $\tau\neq0$. This
indicates that, the sensitivity of the subsystem is equal to the
global one. In noncritical region, the RFS behaves similarly with
the GFS, and $\eta$ depends on $\tau$ but not $N$.

\end{document}